Physical Sciences: Chemistry

# Carbon Nanotubes as Multifunctional Biological Transporters and Near Infrared Agents For Selective Cancer Cell Destruction

Nadine Wong Shi Kam,[1] Michael O'Connell,[1] Jeffrey A. Wisdom,[2] Hongjie Dai[1]*

[1]Department of Chemistry and Laboratory for Advanced Materials, Stanford University, Stanford, CA 94305, USA

[2] Department of Applied Physics, Stanford University, Stanford, CA 94305, USA

* Correspondence to hdai@stanford.edu

Tel: 650-723-4518

Fax: 650-725-0259

**Manuscript Information:** 16 pages of text and 5 figures

Abstract: 160 words

Manuscript: 24,528 characters

**Abstract**

Biological systems are known to be highly transparent to 700-1100nm near-infrared (NIR) light.  It is shown here that the strong optical absorbance of single-walled carbon nanotubes (SWNTs) in this special spectral window, an intrinsic property of SWNTs, can be utilized for optical stimulation of nanotubes inside living cells to afford multifunctional nanotube biological transporters.  For oligonucleotides transported inside living cells by nanotubes, the oligos can translocate into cell nucleus upon endosomal rupture triggered by NIR laser pulses. Continuous NIR radiation can cause cell death due to excessive local heating of SWNT in vitro.  Selective cancer cell destruction can be achieved via functionalization of SWNT with a folate moiety, selective internalization of SWNTs inside cells labeled with folate receptor tumor-markers and NIR triggered cell death, without harming receptor-free normal cells.  Thus, the transporting capabilities of carbon nanotubes combined with suitable functionalization chemistry and their intrinsic optical properties can lead to new classes of novel nanomaterials for drug delivery and cancer therapy.

**Introduction**

The introduction and delivery of DNA, proteins or drug molecules into living cells is important to therapeutics (*1*). Inorganic nanomaterials including nanocrystals, nanotubes and nanowires exhibit advanced physical properties promising for various biological applications (*2-6*) including new molecular transporters. Single-walled carbon nanotubes (SWNTs) have been shown to shuttle various cargos across cellular membrane without cytotoxicity (*7-9*). Much remains to be done to exploit the intrinsic physical properties of SWNTs (*10*) and thus impart unique features to nanotube bio-carriers. Here, we show that while biological systems are transparent to 700-1100nm near-infrared (NIR) light (*11*), the strong absorbance of SWNTs in this window (*12*) can be utilized for optical stimulation of nanotubes inside living cells to afford various useful functions. For oligonucleotides transported inside cells by nanotubes, the oligos can translocate into cell nucleus upon endosomal rupture triggered by NIR laser pulses. Continuous NIR radiation can cause cell death due to excessive local heating of SWNT in vitro. Selective cell destruction can be achieved via functionalization of SWNT with a folate moiety, targeted internalization of SWNTs inside cells labeled with folate receptor tumor-markers (*13*) and NIR triggered cell death, without harming receptor-free normal cells. The intrinsic physical properties of SWNTs can thus be exploited to afford new types of biological transporters with useful functionalities.

**Materials and Methods**

**Water Soluble Cy3-DNA-SWNT Conjugates and Characterization.** As-grown Hipco (*14*) SWNTs were mixed with a 20 µM aqueous solution of Cy3 labeled single-stranded DNA and the initial concentration of SWNTs was ~ 250 mg/L. The DNA sequence was TGGACAAGTGGAATGX, where X denoted the fluorescent label Cy3 and was purchased from the PANfacility at Stanford University. The nanotubes and DNA solution were sonicated for ~ 45 min to 1 hr, and centrifuged at 22, 000 g for ~ 6 hrs. The pellet comprising of impurities, aggregates and bundles of nanotubes at the bottom of the centrifuge tube was discarded and the supernatant was collected and underwent an additional centrifugation round. The resulting supernatant consisted of a solution of SWNT functionalized by Cy3-DNA via non-covalent adsorption (*15*). The solubilized SWNTs were mostly individual tubes (few bundles) as revealed by spectroscopy and microscopy using a Cary 6000i UV-VIS-NIR spectrophotometer and atomic force microscopy (AFM) respectively. The SWNT concentration in the solution after this process was estimated to be ~ 25mg/L (~10% of the starting nanotube suspension). Samples for AFM analysis were prepared by depositing ~ 50 µL of the SWNT suspension on a $SiO_2$ substrate and allowed to stand for 45 min. The substrate was then thoroughly rinsed with distilled water and dried with a $N_2$ stream.

**SWNTs Functionalized by Various Phospholipids (PL).**

**PL with a PEG moiety and Folic Acid Terminal Group (PL-PEG-FA)** 3.5 mM of folic acid (FA, Aldrich) and 5 mM of 1-ethyl-3-(3-dimethylamino-propyl) carbodiimide (EDC, Fluka) were added to a solution of 0.35mM $NH_2$-PEG-PL (2-Distearoyl-*sn*-Glycero-3-Phosphoethanolamine-N-[Amino(Polyethylene Glycol)2000] purchased from Avanti Polar Lipids) in 10 mM phosphate buffer at pH 7.5. After reaction, the solution was dialyzed against phosphate buffer using a membrane (MWCO 1000, Spectrapor) to remove un-reacted FA and EDC. The dialysis was carried out for 3 days with frequent replacement of the buffer. After dialysis, the absorbance of the PL-PEG-FA solution was recorded using a HP-8453 spectrophotometer to ensure that excess free FA was removed from the solution.

**PL with a PEG Moiety and Fluorescein Tag (PL-PEG-FITC).** 3 mg of $NH_2$-PEG-PE was dissolved in 1.5mL of 0.1 M carbonate buffer solution (pH 8.0). To this solution 100 µL of 13mM solution of fluorescein isothiocyanate (FTIC) in DMSO (Aldrich) was added. The mixture was allowed to react overnight at room temperature and protected from light. Purification by gel chromatography was achieved by loading 1 mL of the solution to a Sephadex G-25 column (Aldrich). As elution solvent ($H_2O$) was flown through the column, the formation of two separate yellow bands was observed. The fractions were collected and the absorbance of various fractions was measured at 488 nm using a HP-8453 spectrophotometer. Fractions from the first elution peak were pooled as they were attributed to the higher molecular weight PL-PEG-FITC conjugate (also confirmed by fluorescence measurement), and subsequently used for solubilization of SWNTs.

**Solutions of SWNTs Functionalized with One or Two PL-PEG Molecules (One or Two Cargos).** PL-PEG-FA (one cargo, used later in Fig.5b&5c) or a 1:1 mixture of PL-PEG-FA and PL-PEG-FITC (two cargos, used for Fig. 5d&5e) were used to functionalize and solubilize Hipco SWNTs using the same sonication and centrifuging procedure as described for Cy3-DNA above. The SWNTs functionalized by PL-PEG molecules were also individual tubes characterized by UV-Vis-NIR spectroscopy and AFM.

**Cell Culture and Cellular Incubation in SWNT Solutions.** HeLa cells (an adherent cell line) were cultured in Dubelcco's modified eagle's medium (DMEM) supplemented with 10 % of fetal bovine serum (FBS) and 1 % penicillin-streptomycin (all reagents from Invitrogen). The incubations of Cy3-DNA-SWNT with HeLa cells were carried out (done for Fig. 2,3&4) in 12 well plates, with the cells having been seeded for a period of ~ 18hrs prior to incubation. Cy-DNA-SWNT was added to each well (~ $4x10^5$ cells/well) at a final concentration of ~ 2.5 – 5 mg/L. The incubations were carried out at 37 °C (except for low temperature incubation, where the temperature was 4 °C in Fig. 2c) and in 5% $CO_2$ atmosphere for a period of ~ 12 hrs. After incubation, the cell medium was removed from the well, the cells were washed and detached from the surface by the addition of trypsin-EDTA solution (Invitrogen) for various characterization or laser radiation steps. Similar steps were used for incubating cells in PL-PEG functionalized SWNTs (for Fig.5). Note that all cells were washed with excess SWNTs in the solution removed and placed in fresh solutions after the incubation step and before any of the in-vitro laser radiation experiments described in this work.

**FR[+]Cells and Normal Cells.** HeLa cells were cultured in DMEM medium with folic acid (FA) depleted from the cell medium. It is known that the FA-starved cells over-express folate receptors (FRs) on the cell surfaces . HeLa cells were passaged for at least 4 rounds in the FA-free medium before use to ensure over-expression of FR on the surface of the cells (FR$^+$cells). Normal cells were all cultured in DMEM medium with abundant folic acid and few available free-FRs on the cell surfaces.

**808nm Laser Radiation.** Detached HeLa cells with or without incubation treatment in SWNT solutions were transferred to a circular quartz cuvette (diameter = 3 cm, thickness or optical path length = 1 cm) and exposed to an 808nm laser source, a fiber coupled diode laser bar. Note that all cells were washed with excess SWNTs in the solution removed and placed in fresh solutions after incubation in SWNT solutions and before any of the in-vitro laser radiation experiments described in this work. The diode laser bar was coupled into a 1 meter long 200μm core fiber with a numerical aperture of 0.22. The bare fiber end was imaged to the size of the cuvette using a 50 mm anti-reflection coated bi-convex lens. The center wavelength varied from 806-810 nm depending on the current level, while maintaining a width of 2 nm. A closed loop cooling system suppressed temperature transients to below 0.2˚C to eliminate power and wavelength variations during exposures. Power calibration was performed using a thermal power meter placed after the imaging lens and before the sample. The laser beam size was about 3 cm fully covering the area of the cuvette for radiation of the cells. The power density was tunable up to a maximum of ~ 3.5 W/cm$^2$ and the exposure was carried out either continuously for several minutes or over several 10 sec pulses.

**Confocal Microscopy.** The cells were imaged by a Zeiss LSM 510 confocal microscope. Prior to analysis, the detached HeLa cells (with and without laser exposure) were seeded in chambered coverslides for ~ 12 hrs. For nuclear staining, DRAQ5 (Axxora LLC) was added to each well and allowed to incubate for 5 min at room temperature before confocal imaging.

**Ex-Vitro Measurement of Heating of a SWNT Solution by NIR Radiation.** A DNA-SWNT solution (nanotube concentration 25 mg/L) was irradiated by the 808 nm laser at 1.4 W/cm$^2$. Temperature was measured (Fig. 3c) at 20 s intervals with a thermocouple placed inside the solution for a total period of 2 min. Care was taken to avoid exposure of the thermocouple in the beam path to minimize any direct heating of the thermocouple by the laser. Longer time radiation caused formation of gas bubbles in the solution and eventual boiling of the water solution, as a result of light absorption by SWNTs in the solution. Without nanotubes, an aqueous solution is transparent without heating under the same radiation conditions.

**Cell Proliferation Assay.** CellTiter A96 (Promega), an MTS based assay, was used to monitor cell viability and proliferation after various treatments including internalization of SWNTs and laser radiation. The CellTiter 96 assay uses the tetrazolium compound (3-(4,5-dimethylthiazol-2-yl)-5-(3-carboxymethoxyphenyl)-2-(4-sulfophenyl)-2H-tetrazolium, inner salt; MTS) and the electron coupling reagent, phenazine methosulfate (PMS). MTS is chemically reduced by cells into formazan whose concentration and optical absorbance at 490nm can provide a measure of the metabolically active live cells.

Control cells or cells treated with SWNT and/or laser treatment were left to incubate for a period of up to 15 days at 37°C and 5% $CO_2$ in DMEM. The CellTiter A96 solution, was added to each cell sample and allowed to react for 2 h at 37°C and 5% $CO_2$.

Colorimetric detection and absorbance at 490 nm were used to determine the proliferation profile of different samples.

**Results and Discussion**

The nanotube samples used here were Hipco SWNTs (*14*) solubilized in the aqueous phase by non-covalently adsorbing either 15-mer fluorescently Cy3-labeled single-stranded DNA (*15*) (Fig. 1a) or polyethylene glycol (PEG)-grafted phospholipids (PL-PEG) (Fig. 5a). These nanotube solutions were highly stable in buffer solutions (Fig. 1c insert), consisting of very pure, short (average length~150nm) and individualized SWNTs (by sonication and centrifugation (*12*) rather than bundles as evidenced by UV-Vis-NIR absorbance (Fig. 1b) and atomic force microscopy (AFM, Fig. 1d) data. The molar extinction coefficient of the solubilized SWNTs (molecular weight MW~170kD for length~150nm, diameter ~ 1.2nm) measured at $\lambda$=808nm in the NIR was $\varepsilon \sim 7.9 \times 10^6$ $M^{-1}cm^{-1}$ (Fig. 1c). The high absorbance of SWNTs in the NIR originates from electronic transitions between the first or second van Hove singularities of the nanotubes (*12,16*) High optical absorbance of SWNTs in the 700-1100nm NIR window transparent to biological systems (*11*) is exploited in the current work at a single wavelength by using a $\lambda$=808 nm laser (beam size ~ 3cm and power density up to 3.5W/cm$^2$) for in-vitro radiation.

By confocal fluorescence microscopy imaging, we observed that upon exposure of HeLa cells to a Cy3-DNA-SWNT solution at 37 °C, the DNA-SWNT conjugates were internalized (Fig. 2a&2b) inside the cells with nanotubes as the transporters. The green color in Fig. 2a corresponds to Cy3 labels on DNA-SWNTs inside HeLa cells. After staining the cell nucleus with a DRAQ5 dye, we carried out dual color detection and observed accumulation of DNA-SWNT in the cytoplasm region with little colocalization of Cy3-DNA in the nucleus (Fig. 2b). This suggested lack of nuclear translocation for the DNA molecules carried across the cell membranes by nanotubes. Experiments carried out at 4°C found no uptake of Cy3-DNA-SWNT conjugates inside cells (Fig. 2c), suggesting an energy-dependent endocytosis mechanism (*17*) for the uptake observed at 37°C.

Endocytosis is known to rely on enclosure of molecules inside endosomes or lipid vesicles during and after cell entry (*17*). Motivated by the need of endosomal rupture for efficient molecular releasing and delivery (*18*), we explore the effects of NIR light on DNA-SWNTs after endocytosis. We first note that control experiments find that cells without nanotubes are highly transparent to NIR and exhibit no ill effect after radiation for up to 5 min by a 3.5 W/cm$^2$ $\lambda$=808 nm coherent laser light (laser beam diameter ~3 cm uniformly radiating over the entire area of the cuvette containing the cells). For HeLa cells after DNA-SWNT uptake, we have experimented with the NIR radiation conditions and found that six 10s on-and-off pulses of 1.4 W/cm$^2$ laser radiation can afford releasing effects without causing cell death. After such treatment, confocal imaging reveals colocalization of fluorescence of Cy3-DNA in the cell nucleus (Fig. 3a), indicating releasing of DNA cargos from SWNT transporters and nuclear translocation after the laser pulses.

To glean the effects of NIR optical excitation of SWNTs inside living cells, we carried out a control experiment by radiating an aqueous solution of Cy3-DNA functionalized SWNTs ex-vitro. We observed that radiation of a SWNT aqueous solution (nanotube concentration 25mg/L) by 1.4W/cm$^2$ λ= 808 nm laser continuously for 2 min caused heating of the solution to ~70 ºC (Fig. 3c, boiling of solution was observed for even longer radiations). Without solubilized nanotubes, the solution was transparent to 808 nm light with little heating detected. These clearly showed that optically stimulated electronic excitations of SWNTs rapidly transferred to molecular vibration energies and caused heating. Another phenomenon was that after ex-vitro NIR radiation of a Cy3-DNA/SWNT solution (without cells), an apparent increase in the Cy3 fluorescence was observed (Fig. 3b), indicating unwrapping and releasing of Cy3-DNA strands from nanotubes and thus reduced quenching of Cy3 by nanotubes. Taken together, the results suggest that SWNTs internalized in living cells can act as tiny NIR 'heaters' or 'antennas'. Opto-electronic excitations of nanotubes inside cells by NIR radiation can trigger endosomal rupture and releasing of non-covalent molecular cargos from nanotube carriers. Once detached from nanotubes and freed into the cytoplasm, the DNA molecules diffuse freely across the nuclear membrane (*19*) into the nucleus.

No apparent adverse toxicity effects were observed with cells after SWNT endocytosis and NIR-pulse (1.4W/cm$^2$) activated DNA releasing and nuclear translocation in terms of short term viability (Fig. 4a and Supp. Info.) and long term cell proliferation (Supp. Info). In a control experiment, cells without exposure to SWNTs survived continuous 3.5W/cm$^2$ 808 nm laser radiation for 5 min (Fig.4b), clearly illustrating high transparency of bio-systems to NIR light. In stark contrast, for cells with internalized SWNTs, extensive cell death was observed after 2 min radiation under a 1.4W/cm$^2$ power as evidenced by cell morphology changes (Fig. 4c vs. Fig. 4a&4b), loss of adherence to substrates and aggregation of cell debris (Fig. 4c). Extensive local heating of SWNTs inside living cells due to continuous NIR absorption was the most likely origin of cell death. Interestingly, we observed that the dead cells 'released' SWNTs to form black aggregates floating in the cell-medium solution visible to the naked eye approximately 24 h following irradiation (Fig. 4e inset, black specks). Raman spectroscopy and scanning electron microscopy (SEM) identified SWNTs mixed with cell debris in the black aggregates. The $266cm^{-1}$ Raman signal corresponded to aggregated SWNT bundles (*20*) while a broad photoluminescence peak observed around $3200 \text{ cm}^{-1}$ (~1050 nm) (Fig.4c) corresponded to individual tubes still in existence. SEM of the black aggregates after drying revealed tube-like strands stretched across cell debris or residues from the cell culture medium (Fig. 4d, inset). Cracks appeared in the nanotube-cell debris structures during drying, causing the aggregated bundles of nanotube to stretch across the cracks. Thus, we clearly observed that accompanied by cell death, extensive NIR radiation caused molecular-detachment or de-functionalization of SWNTs inside cells leading to nanotube aggregation.

The result above hinted that if SWNTs can be selectively internalized into cancer cells with specific tumor markers, NIR radiation of the nanotubes in-vitro can then selectively activate or trigger cell death without harming normal cells. This important goal prompted us to develop SWNT functionalization schemes with specific ligands for recognizing and targeting tumorous cell types. Folate receptors (FR) are common tumor markers expressed at high levels on the surfaces of various cancer cells and facilitate cellular internalization of folate-containing species via receptor-mediated endocytosis (*13*). To exploit this system, we obtained highly water soluble individualized SWNTs non-covalently functionalized by phospholipids PL-PEG-folic acid (PL-PEG-FA) (Fig.5a). FR-positive HeLa cells (FR$^+$cells) with over-expressed FRs on the cell surfaces were obtained by culturing cells in folic acid-depleted cell medium. Both FR$^+$cells and normal-cells without abundant free FRs were exposed to (PL-PEG-FA)-SWNTs for 12-18h, washed and then radiated by 808 nm laser (1.4 W/cm$^2$) continuously for 2 min. After the NIR radiation, we observed extensive cell death for the FR$^+$cells evidenced by drastic cell morphology changes (Fig. 5b) while the normal-cells remained intact (Fig. 5c) and exhibited normal proliferation behavior over ~ 2 weeks that was the longest period monitored. The selective destruction of FR$^+$cells suggested that (PL-PEG-FA)-SWNTs were efficiently internalized inside FR$^+$cells (confirmed by fluorescence in Fig.5d for SWNTs with FA cargo and FITC labels) and not inside normal-cells (confirmed by the lack of fluorescence inside cells in Fig.5e). The former was a result of selective binding of FA-functionalized SWNTs and FRs on FR$^+$cell surfaces and receptor-mediated endocytosis. The latter was due to the 'inertness' or blocking of non-specific binding of SWNTs imparted by the PEG moiety (*6*) on SWNTs and the lack of available FRs on the normal cells.

It is shown here that single walled carbon nanotubes are molecular transporters or carriers with very high optical absorbance in the near-infrared regime where biological systems are transparent. This intrinsic property stems from the electronic band structures of nanotubes and is unique among transporters. Our current work exploits this property with a $\lambda$=808 nm laser and can be extended to using light sources spanning the entire 700-1100 nm range transparent to biosystems for more efficient in-vitro excitations of SWNTs with various chiralities to obtain enhanced biological effects. NIR pulses can induce local heating of SWNTs in vitro for endosomal rupture and molecular cargo releasing for reaching intended targets without harming cells. On the other hand, selective killing of cells over-expressing tumor markers can be achieved by selective delivery of nanotubes inside the cells via receptor-mediated uptake pathways and NIR triggered death. The scheme of SWNT functionalization by PEG-ligands can be generalized to various ligands or antibodies targeting very specific types of cells. While the PEG-moiety imparts inertness and little non-specific binding of nanotubes to normal cells, the ligands can recognize cells with complementary receptors for SWNT internalization and subsequent cell destruction by NIR radiation. Specifically functionalized nanotubes could then be a generic 'killer' of various types of cancer cells without harming normal cells. Thus, the transporting capabilities of carbon nanotubes combined with suitable functionalization chemistry and the intrinsic optical properties of SWNTs can open up exciting new venues for drug delivery and cancer therapy.

**Acknowledgments.** We are grateful to Dunwei Wang for SEM analysis. This work was supported by the Stanford NSF CPIMA.

**Figure Captions:**

**Figure 1.** Carbon nanotubes with high NIR absorbance solubilized in water. (a) schematic of a Cy3-DNA functionalized SWNT. (b) UV-Vis spectra of solutions of individual-SWNTs functionalized non-covalently by 15-mer Cy3 labeled-DNA at various nanotube concentrations (top curve: SWNT concentration ~ 25mg/L in $H_2O$, lower curves correspond to consecutive 3% reduction in SWNT concentration. The well-defined peaks in the UV-Vis spectra suggest individual SWNTs in the solutions obtained by removing bundles via centrifugation. (c) Absorbance (A) at 808 nm vs. SWNT concentration (optical path = 1cm). Solid line is Beer's law fit to obtain molar extinction coefficient of SWNT $\varepsilon \sim 7.9 \times 10^6$ $M^{-1}$ $cm^{-1}$. Inset: a photo of a DNA functionalized SWNT solution. (d) Atomic force microscopy (AFM) image of DNA functionalized individual SWNTs (height 1-1.5nm) deposited on a $SiO_2$ substrate, scale bar = 200 nm.

**Figure 2.** Transporting DNA inside living cells by SWNTs. (a) A confocal fluorescence image (excitation $\lambda$=548 nm, emission detected at $\lambda$=560 nm) showing the internalization and accumulation of Cy3-DNA-SWNT around the nucleus (circular regions surrounded by green fluorescence corresponding to Cy3) of HeLa cells after incubation of cells (~ $4 \times 10^5$ cells/well in 12-well plates) for 12 h at 37°C in a 2.5 to 5 mg/L Cy3-DNA-SWNT solution . (b) Dual detection of Cy3-DNA-SWNT (green) internalized into a HeLa cell with the nucleus stained by DRAQ5 (red). (c) A confocal image of HeLa cells after incubation in a Cy3-DNA-SWNT solution at a low temperature of 4°C. Only DRAQ5 stained nucleus (red color) of HeLa is seen. The lack of green fluorescence detected indicates that there is minimal cellular uptake of the Cy3-DNA-SWNT conjugates at the low temperature.

**Figure 3.** In-vitro near-infrared excitation of SWNT transporters for DNA cargo releasing and nuclear translocation. (a) A confocal image of HeLa cells after 12h incubation in a 2.5 to 5 mg/L Cy3-DNA-SWNT solution for internalization and radiation by 6 NIR (808 nm) 10s-pulses (@1.4W/$cm^2$ power density). Colocalization (yellow color) of Cy3-DNA (green) in cell nucleus (red) was detected indicating translocation of Cy3-DNA to the nucleus. After incubation, the cells were washed and re-suspended in cell medium in a quartz cuvette for NIR radiation (laser beam diameter ~ 3cm, power 10W, optical path 1cm).  (b) Cy3 fluorescence emission spectra of a Cy3-DNA-SWNT solution (25 mg/L) before (blue curve) and after (red curve) laser radiation (1.4 W/cm$^2$) for 2 min. $\lambda_{excitation}$ = 550 nm and $\lambda_{emission}$ = 563 nm.  (c) An ex-vitro control experiment. Temperature evolution of a DNA-SWNT solution (~25 mg/L) during continuous radiation by a 808nm laser @1.4 W/cm$^2$ for 2 min. This result clearly reveals heating of solution due to absorption of 808 nm laser light by SWNTs in the solution.

**Figure 4.**  Fate of cells with internalized SWNTs after NIR laser pulses or extensive radiations. (a) Optical image of HeLa cells after DNA-SWNT uptake and 6 pulses of 10s-long 808nm laser radiations @1.4 W/cm$^2$. Cells retained normal morphology with no apparent death observed. Inset: photo of the cell solution taken 12 h after laser pulses. Pink color due to phenol red in the well. HeLa cells remained adhered to the bottom of the container. (b) Image of HeLa cells without internalized SWNTs after continuous 808nm laser radiation @3.5 W/ cm$^2$ for 5 min.  No cell death was observed. (c) Image of dead and aggregated cells after internalization of DNA-SWNT and laser radiation @1.4 W/cm$^2$ for 2 min.  The dead cells showed rounded and aggregated morphology as opposed to live cells in a 'stretched' form in (a)&(b). Inset: photo of the cell solution 24 h after laser activated cell death. No live cells adherent to the bottom of the container was observed. Black aggregates containing SWNTs released from dead cells floating on water were visible (indicated by arrow) to the naked eye. (d) Raman data and SEM image (inset) of the black aggregates after drying.

**Figure 5.**  Selective targeting and killing of cancer cells. (a) Chemical structure of (PL-PEG-FA) and (PL-PEG-FITC) synthesized by conjugating PL-PEG-NH2 with folic-acid or FITC respectively for solubilizing individual SWNTs. (b) Top panel: schematic of selective internalization of (PL-PEG-FA)-SWNTs into folate-over-expressing (FR$^+$) cells via receptor binding and then undergoing NIR 808 nm laser radiation. Lower panel: image showing death of FR$^+$cells with rounded cell morphology after the process in the top panel (808nm laser radiation at 1.4W/cm$^2$ for 2 min). Higher magnification image (inset) shows details of the killed cells. (c) Top panel: schematic of no internalization of (PL-PEG-FA)-SWNTs into normal cells without available FRs. Lower panel: image showing normal cells with no internalized SWNTs are unharmed by the same laser radiation condition as in (b). Higher magnification image (inset) shows a live normal cell in 'stretched' shape. (d) Confocal image of FR$^+$cells after incubation in a solution of SWNTs with two cargos, (PL-PEG-FA) and (PL-PEG-FITC). The strong green FITC fluorescence inside cells confirms the SWNT uptake with FA and FITC cargos. (e) Same as (d) for normal cells without abundant FRs on cell surfaces. There is little green fluorescence inside cells confirming little uptake of SWNTs with FA and FITC cargos.

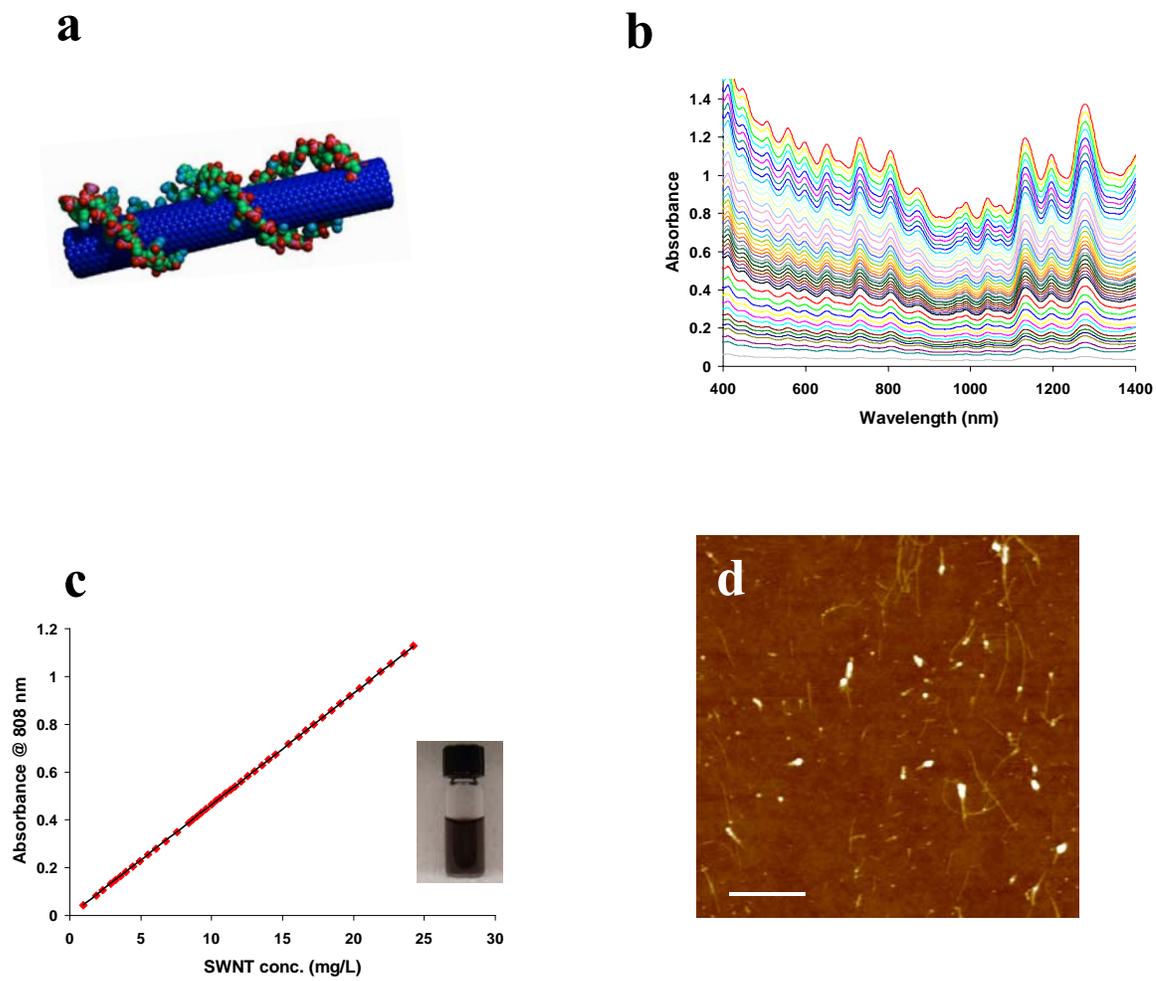

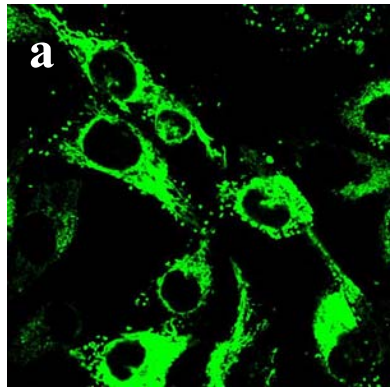

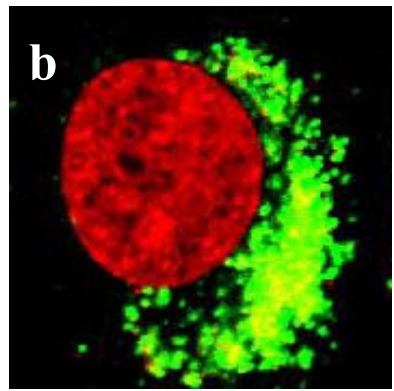

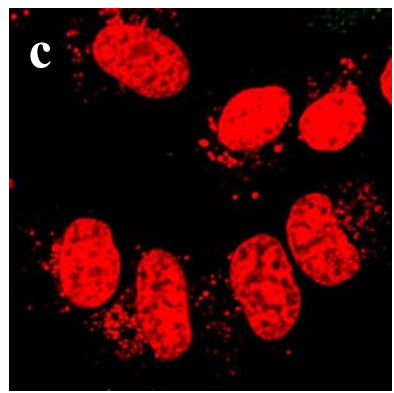

Figure 2

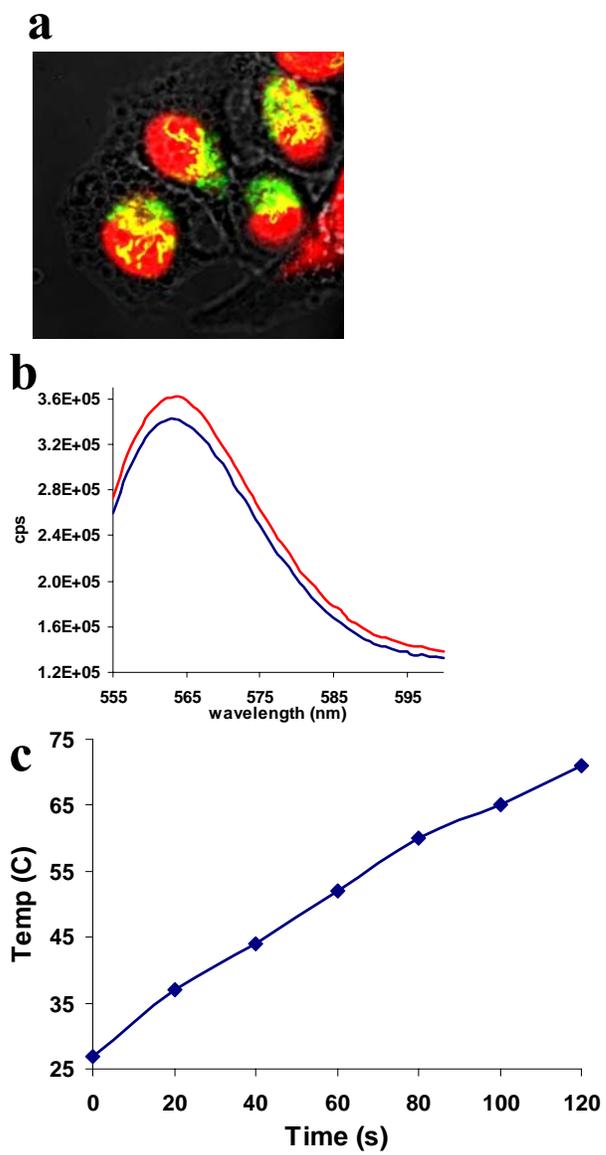

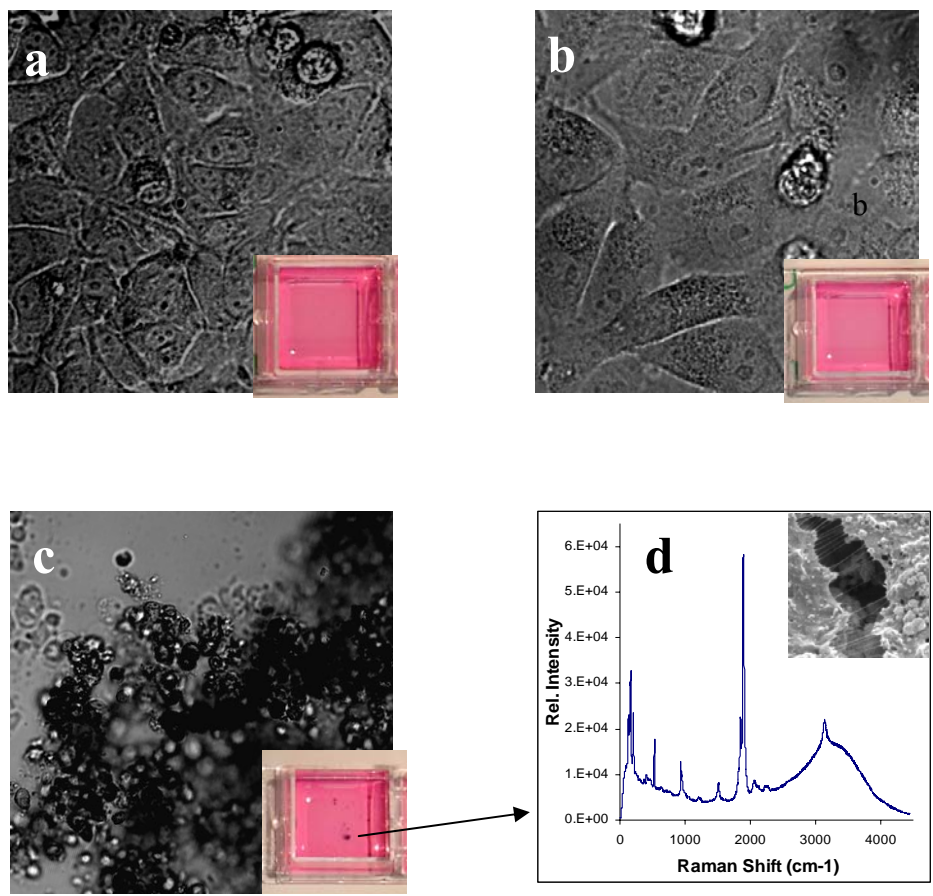

Figure 5